
\input phyzzx
\PHYSREV
\hoffset=0.3in
\voffset=-1pt
\baselineskip = 14pt \lineskiplimit = 1pt
\frontpagetrue
\rightline {Cincinnati preprint March.1995, RU-2-95}
\medskip
\titlestyle{\seventeenrm Asymptotic Freedom in a String Model of High
Temperature QCD}
\vskip.5in
\medskip
\centerline {\caps M. Awada\footnote*{\rm E-Mail address:
moustafa@physunc.phy.uc.edu}}
\centerline {Physics Department}
\centerline {\it University of Cincinnati,Cincinnati, OH 45221}
\centerline {\it and}
\centerline {High Energy Theory Group}
\centerline {\it The Rockefeller University, New York, NY 10021}

\bigskip
\centerline {\bf Abstract}
\bigskip
Recently we have shown that a phase transition occurs
in the leading and sub-leading approximation of the large N limit in
rigid strings coupled to long range  Kalb-Ramond interactions.
The disordered phase is essentially the Nambu-Goto-Polyakov string theory
while the ordered phase is a new theory.  In this letter we compute the free
energy per unit length of the interacting rigid string at finite temperature.
We show that the mass of the winding states solves that of QCD strings in
thelimit of high temperature.  We obtain a precise identification of the QCD
couplingconstant and those of the interacting rigid string.  The relation we
obtain
is $Ng_{QCD}^2 = ({4\pi^2 (D-2)\over 3})^2{1\over 3\kappa}$ where
$\kappa = {D t \alpha\over \pi \mu_{c}}$ is the ratio of the extrinsic
curvature coupling constant t, the Kalb-Ramond coupling constant $\alpha$,
and the critical string tension $\mu_{c}$.  The running beta function of
$\kappa$ reproduces correctly the asymptotic behaviour of QCD.

\eject

Polyakov [1] has argued that the string theory appropriate to QCD should
be one with long range correlations of the unit normal. The Nambu-Goto
(NG) string theory is not the correct candidate for large N QCD as it
disagrees with it at short distances.  The NG string does not give
rise to the parton-like behaviour observed in deep inelastic scattering
at very high energies.  The observed scattering amplitudes have a power
fall-off behaviour contrary to the exponential fall-off behaviour of
the NG string scattering amplitudes at short distances.  The absence of
scale and the power law behaviour at short distances suggest that
the QCD string must have long range order at very high energies.
Pursuing this end, Polyakov considered modifying the Nambu action
by the renormalizable scale invariant curvature squared term
(rigid strings).  The theory closely resembles the two dimensional
sigma model where the unit normals correspond to the sigma fields.
In the large N approximation, there is no phase transition.
In fact the high temperature limit of the free energy of the free
rigid string is inconsistent (being an imaginary quantity) and disagrees
with the QCD strings [2].  However there is a dimensional agreement
which suggests that the free rigid string has some relevance to QCD.
The absence of phase transition to a long range order in the free
rigid string and the disagreement with QCD at finite high temperature
past the transition point are correlated.  Therefore, it was crucial
to first construct a string theory with long range order and then address
the questions of [2].  In [3], and [4] we coupled the rigid strings
to long range Kalb-Ramond fields.  Since spin systems in two dimensions may
exhibit a phase transition with the inclusion of long range interactions,
it is natural to conjecture likewise for rigid strings
with long range Kalb-Ramond fields.  Indeed  we proved that there is
a phase transition to a region of long range order in
the leading and sub-leading order of large N approximation where N is
the space-time dimensions.  Such a theory may therefore be relevant to QCD.

In this letter we compute the free energy per unit
length of the interacting rigid string at finite temperature.  We show that
the mass of the winding states not only has the correct dimensions,
but also the correct sign as that of QCD strings in the limit of high
temperature.  Consequently
we obtain a precise identification of the QCD coupling constant and those of
the interacting rigidstring.  The relation we obtain is $Ng_{QCD}^2 = ({4\pi^2
(D-2)\over 3})^2{1\over 3\kappa}$ where
$\kappa = {D t \alpha\over \pi \mu_{c}}$ is the ratio of the extrinsic
curvature coupling constant t, the Kalb-Ramond coupling constant $\alpha$,
and the critical string tension $\mu_{c}$.  The running beta function of
$\kappa$ reproduces correctly the asymptotic behaviour of QCD.
We remark that the QCD formula that we compare with is that of Polchinski
[5] who claims that it describes the high temperature limit of the confining
phase of QCD.  However the results of [6] shows that at large N there is no
analytical
continuation of the low temperature confining phase past the
transition to the arbitrarily high temperature deconfining phase.
The calculations of [6] show branch cut singularities in the Wilson-two
point function in the deconfining phase signaling the propogation of
quarks and gluons.  Provided that the results of [6] persist in higher
loop corrections, one should interpret the results of [5] and [7] in
the high temperature deconfining phase.

 The gauge fixed action of the rigid string [3,4] coupled to the rank
two antisymmetric Kalb-Ramond tensor field $\phi$ [8] is :
$$S_{gauge-fixed} = \mu_{0}\int d^2\xi\rho + {1\over 2t_{0}}\int d^2\xi
[\rho^{-1}(\partial^{2}x)^{2} + \lambda^{ab}(\partial_{a}x
\partial_{b}x - \rho\delta_{ab})] + S_{K-R}\eqno{(1 a)}$$
where
$$S_{K-R}= e_{0}\int d^2\xi \epsilon^{ab}\partial_{a}x^{\mu}
\partial_{b}x^{\nu}\phi_{\mu\nu} + {1\over 12}\int d^4x
F_{\mu\nu\rho}F^{\mu\nu\rho}\  .\eqno{(1 b)}$$
where $e_{0}$ is a coupling constant of dimension $length^{-1}$,
$t_{0}$ is the bare curvature coupling constant which is dimensionless
and F is the abelian field strength of $\phi$.  The integration of the
$\phi$ field is Gaussian.  We obtain the following interacting
long range Coulomb-like term that modifies the rigid string:
$${1\over 2t_{0}}\int\int d^2\xi d^2\xi' \sigma^{\mu\nu}(\xi)
\sigma_{\mu\nu}(\xi')V(|x-x'|, a)\eqno{(1 c)}$$
where V is the  analogue of the long range Coulomb potential:
$$ V(|x-x'|,a)= {2g_{0}\over \pi}{1\over |x(\xi)-x(\xi')|^{2}
+a^{2}\rho}\   .\eqno{(1 d)}$$
where $\sigma^{\mu\nu}(\xi)= \epsilon^{ab}\partial_{a}x^{\mu}
\partial_{b}x^{\nu}$.
We have introduced the cut-off "a" to avoid the singularity at
$\xi=\xi'$ and define $ g_{0}=t_{0}\alpha_{Coulomb}=t_{0}
{e_{0}^{2}\over 4\pi}$ which has dimension of $length^{-2}$.
The partition function is
$$ Z =\int D\lambda D\rho Dx exp(-S_{eff})\  .\eqno{(2)}$$
where the effective action $S_{eff}$ is (1a) and (1c).
We will assume that $t_{0}$ is small for the sake of obtaining analytical
results. We will evaluate the partition function in a periodic
Euclidean space-time (this would be equivalent to the partition function
at finite temperature).  Thus we consider the world sheet to be
annulus, with a modulus $\tau$ such that the $\xi_{2}$ runs from
0 to $\beta\tau$ where $\beta$ is the inverse temperature and $\xi_{1}$
runs from 0 to L the length of the path.
We will expand around  the configuration:
$$ x_{0}^{1}(\xi)=\xi_{1},~~~~x_{0}^{2}(\xi)=
{\xi_{2}\over \tau}\eqno{(3 a)}$$
$$ \rho(\xi) = \rho_{0},~~~~~~~~\lambda^{ab}(\xi)=\lambda_{0}^{ab}
\eqno{(3 b)}$$

The long range KR interactions are non-local and impossible to
integrate.  Therefore we consider
$$x^{\mu}(\xi) = x^{\mu}_{0}(\xi) + x^{\mu}_{1}(\xi)$$
and expand the Coulomb term (1c,d) to quadratic order in
$x^{\mu}_{1}(\xi)$ about the background straight line $x_{0}$.
The x-integration is now Gaussian and to the leading D approximation
we obtain the new effective action $S_{eff}$:
$${S}_{eff} = S_{0} + S_{1}$$
where
$$S_{0} =  {L\beta\tau\over 2t_{0}}[\lambda_{0}^{11}
+\tau^2\lambda_{0}^{22} +
\rho_{0}(2t_{0}\mu_{0}-\lambda^{ab}\delta_{ab} +8g )]\eqno{(4 a)}$$
$$ S_{1} = {(D-2)\over 2}trln A\eqno{(4 b)}$$
and A is the operator
$$ A = \partial^{2}\rho^{-1}\partial^{2} -\partial_{a}
\lambda^{ab}\partial_{b} + V(\xi,\xi')\  .\eqno{(4 c)}$$
In computing $S_{0}$ we encountered a classical renormalization of the
K-R coupling constant $g_{0}$ and the renormalized coupling constant
is defined by
$$ g = g_{0}ln{\beta^2\tau^2\over a^2}\  .\eqno{(4 d)}$$
As we will see later this formula appears precisely in the computation of
the one loop effective action $S_{1}$ whose explicit form is given by:
$$S_{1} = {(D-2)L\over 2}\sum_{n=-\infty}^{+\infty}\int_{-\infty}^
{+\infty}{dk_{1}\over 2\pi} lnG^{-1}(k_{1},n)$$
$$G^{-1}(k_{1},n)=(p^4 +\rho_{0}[k_{1}^2\lambda_{0}^{11} +
k_{n}^2\lambda_{0}^{22} + p^2 V_{0}(|p|) + V_{1}(|p|)])
\eqno{(5 a)}$$
where
$$ V_{0}(p) = {4g_{0}\over \pi}\int d^2\xi
{e^{ip.\xi}\over \xi^{2}+a^{2}}= 8g_{0}K_{0}(a|p|)\eqno{(5 b)}$$
$$V_{1}(p)={8g_{0}\over \pi}\int d^2\xi
{[e^{ip.\xi}-1]\over (\xi^{2}+a^{2})^{2}}= {8g_{0}\over a^2}
(a|p|K_{1}(a|p|)-1)\eqno{(5 c)}$$
where $K_{n}(z), n=0,1,..$ is the Bessel function of the third
kind,
$$K_{1}(z) := -{d\over dz}K_{0}(z)\  .\eqno{(6)}$$
and
$$ p^2 = (k_{1}^2 +k_{n}^2) ;~~~k_{n}^2 =4\pi^2
n^2\beta^{-2}\tau^{-2}\eqno{(7)}$$
In the large D limit the stationary point equations resulting
from varying with respect to $\lambda_{0}^{ab}$, $\rho$, and $\tau$
respectively are:
$$\rho_{0}^{-1}= 1 - {(D-2)t_{0}\over \beta \tau}I_{0}\eqno{(8 a)}$$
$$\rho_{0}^{-1}\tau^{-2} =1 - {(D-2)t_{0}\over \beta \tau}I_{1}
\eqno{(8 b)}$$
$$\lambda_{0}^{22} =(2t_{0}\mu_{0}-\lambda^{11} +8g )+
{(D-2)t_{0}\over \beta \tau}(\lambda_{0}^{11}I_{0} +
\lambda_{0}^{22}I_{1} + 8g_{0}I_{2})\eqno{(8 c)}$$
$$\lambda_{0}^{11}=(t_{0}\mu_{0} +4g) + {3(D-2)t_{0}\over 2\beta
\tau}(\lambda_{0}^{11}I_{0} +\lambda_{0}^{22}I_{1})+
{(D-2)t_{0}\over \beta \tau}(\rho_{0}^{-1}I_{4} +
8g_{0}[I_{2}+I_{3}])\eqno{(8 d)}$$
where the I's are integrals of the following forms:
$$I_{0}= \sum_{n=-\infty}^{+\infty}\int_{-\infty}^
{+\infty}{dk_{1}\over 2\pi}k_{1}^{2}G(k_{1},n)\eqno{(9 a)}$$
$$I_{1} =\sum_{n=-\infty}^{+\infty}k_{n}^2\int_{-\infty}^
{+\infty}{dk_{1}\over 2\pi}G(k_{1},n)\eqno{(9 b)}$$
$$I_{2} =\sum_{n=-\infty}^{+\infty}\int_{-\infty}^
{+\infty}{dk_{1}\over 2\pi}{V(a|p|)\over a^2}G(k_{1},n)\eqno{(9 c)}$$
$$I_{3} ={1\over 2}\sum_{n=-\infty}^{+\infty}k_{n}^2\int_{-\infty}^
{+\infty}{dk_{1}\over 2\pi}[K_{0}(a|p|)-a|p|K_{1}(a|p|)]
G(k_{1},n)\eqno{(9 d)}$$
$$I_{4} =2\sum_{n=-\infty}^{+\infty}k_{n}^2\int_{-\infty}^
{+\infty}{dk_{1}\over 2\pi}p^2G(k_{1},n)\  .\eqno{(9 e)}$$
The potential $V(a|p|)$ is defined to be
$$V(a|p|) = a^2p^2K_{0}(a|p|)+a|p|K_{1}(a|p|)-1\  .\eqno{(10)}$$
To calculate the integrals (9) exactly is an impossible task, however
our interest is only in the high temperature limit.  In this limit and
in the continuum limit $a\rightarrow 0$ we evaluate the integrals in two
steps: first we evaluate the contributions at n=0, and then for $n>0$
we neglect the quadratic term in $k_{1}^2$ (second derivative term in
(4b,c)) compared to the quartic term $k_{1}^4$ (fourth derivative term
in (4b,c)).  The high temperature limit is regarded when $S_{1}>>S_{0}$
i.e $\beta^{-2} >> \mu_{0} + 4{g_{0}\over t_{0}}$.  The values of the
integrals in (9) are:
$$I_{0} = {Z\over 2}[Z\rho_{0}(\lambda_{0}^{11} +2g)]^{-{1\over 2}}
+ O(\beta)\eqno{(11 a)}$$
$$I_{1} = 0+O(\beta)\eqno{(11 b)}$$
$$I_{2} = {1\over 4}ln{\beta^2\tau^2\over a^2}I_{0}
-Z({3\over 64}a^2ln{\beta^2\tau^2\over a^2})
[Z\rho_{0}(\lambda_{0}^{11} +2g)]^{{1\over 2}}+ O(\beta)\eqno{(11 c)}$$
$$I_{3} = Z({3\over 32}a^2ln{\beta^2\tau^2\over a^2})
(-{\pi\over 3\beta\tau})+ O(\beta)\eqno{(11 d)}$$
$$I_{4} =Z( -{\pi\over 3\beta\tau})+ O(\beta)\  .\eqno{(11 e)}$$
where g is the renormalized coupling constant as given in (4d) and
$$Z^{-1} = 1 + {3\over 32}(8g_{0}a^2\rho_{0})ln{\beta^2\tau^2\over a^2}
\  .\eqno{(12)}$$
The phase transition of the theory of interacting rigid string
with the K-R fields is characterized by two dimensionless coupling
constants, the extrinsic curvature coupling constant $t_{0}$ and
another one defined by [3,4] $\eta_{0}:=8g_{0}a^2\rho_{0}
:= \kappa_{0}\rho_{0}$ appearing in the
critical line equation of the theory at zero temperature.
In [4] we have proved that $\eta_{0}$ is a positive function
of ${Dg_{0}\over \pi\mu_{c}}$, in particular at very
high energies one finds $\rho_{0} =1$, and
$\kappa_{0} = {Dg_{0}\over \pi\mu_{c}}$
where $\mu_{c}$ is the value of the string tension on the critical line.
Owing to the definition of $g_{0}$ which is $t_{0}\alpha$ in (1d), then
$$\kappa_{0} = {Dt_{0}\alpha_{0}\over \pi\mu_{c}}\  .\eqno{(13)}$$
is a remarkable combination of the three coupling constants of the theory.
It follows from the definition of $\kappa_{0}$, the stationary
point equations (8), and the results (11) that we can define
a renormalized coupling $\kappa$ as:
$$\kappa(\Lambda={1\over a}) = \kappa_{0}ln{\beta^2\tau^2\over a^2}\
.\eqno{(14)}$$
This is one of the main crucial results of this paper.  Equation (14)
will play a very important role in obtaining correctly the asymptotic
behaviour of QCD.  The running beta function is therefore:
$$\beta(({\kappa\over \kappa_{0}})^{-1}) =
-({\kappa\over \kappa_{0}})^{-2}
\footnote*{We can set $\kappa_{0} \sim 1$ since $g_{0} \sim {1\over a^2}$}
\  .\eqno{(15)}$$
 From (8b) we deduce that:
$$\rho_{0}^{-1}=\tau^{2}\eqno{(16 a)}$$
and from (8a) we obtain in the high temperature limit:
$$\lambda_{0}^{11} +2g = {1\over (1-\tau^{2})^2}
(1+{\epsilon\over \tau^2})^{-1}{(D-2)^2\over 4}
{t_{0}^2\over \beta^2}\  .\eqno{(16 b)}$$
Using (16b) we can completely determine $\lambda_{0}^{22}$ from (8c)
in terms of $\tau$,$\beta$, and $t_{0}$. Equation (8d) determines $\tau$
in terms of the couplings of the theory.  We obtain the
following $\tau$ equation:
$$ \delta({1\over (1-\tau^{2})^2}-{3\over 2} {1\over (1-\tau^{2})}
+\epsilon{1\over \tau^2(1-\tau^2)}(1+{\epsilon\over \tau^2})^{-1})
 +(1+{\epsilon\over \tau^2})=0\eqno{(17)}$$
where
$$\epsilon = {3\over 32}\kappa~~;~~\delta={3(D-2)t_{0}\over 4\pi}
\  .\eqno{(18)}$$

In the absence of the long range K-R interactions ($\epsilon=0$)
eq.(17) becomes a quadratic equation whose solutions are the complex
valued solution of Polchinski [2]:
$$\tau^2 = 1 \pm i(\delta)^{1\over 2}\eqno{(19)}$$
leading to an inconsistent result
in the sense that the mass square of the winding states of the free
rigid string is imaginary.

The presence of the K-R interactions critically changes the situation and
equation (17) can be rewritten as a quartic equation which has two real
solutions in terms of the square of the modulus $\tau$ and
two complex solutions:
$$\tau_{1}^{2} = -\epsilon(1 \pm ({\delta\over 1+\epsilon})^{1\over 2}
+O(\delta))\eqno{(20)}$$
$$\tau_{2}^{2}=1 \pm i({\delta\over 1+\epsilon})^{1\over 2}
 + O(\delta)\  .\eqno{(21)}$$
In fact we will show that these complex solutions are not good
saddle points of the effective action in the sense that they do not
give the best estimate, the maximum value, of the effective action.
To determine the effective action we integrate eq.(8) and use the results
(11),(12), and (14) together with the saddle point solutions for
$\rho_{0}$ and $\lambda_{0}^{11}$ in (16).  We express the result
in terms of the square mass $M_{1, int.rigid}^2$ of the winding states
whose definition is $M_{1, int.rigid}:={S_{eff}\over L}$.
We obtain
$$M_{1,int.rigid}^2 = ({\pi(D-2)\over 3})^2{1\over \beta^2\tau^2}
(1 - {\delta\over 2(1-\tau^2)}(1+{\epsilon\over \tau^2})^{-1})^2
\  .\eqno{(22 a)}$$
The inverse power dependence on $\beta^2$ is a consequence of the fact
that the UV behaviour of our model is governed by dimensionless couplings.
One can easily check that the $\tau$ equation (17) is the stationary
equation of the effective action $S_{eff}(\tau,\epsilon,\delta)$.
Indeed (17) is nothing but
$${\partial\over \partial \tau}M_{1, int.rigid}(\tau) = 0
\  .\eqno{(22 b)}$$
Substituting (19) into (22a), it is straightforward to obtain the result
in [2] in the absence of the K-R long range interactions, $\epsilon = 0$.
However the saddle point solutions (21) do not correspond to the maximum
of the effective action.  Using the method of steepest descent and
Cauchy's theorem, a good saddle point solution $\tau=\tau^*$ must satisfy:
$$Re({\partial^2\over \partial \tau^2}S_{eff}(\tau))|_{\tau=\tau^*} < 0
\  .\eqno{(23)}$$
One can easily check that (21) does not satisfy (23) and
$Re({\partial^2\over \partial \tau^2}S_{eff}(\tau))|_{\tau=\tau_{2}}$
is actually positive.  Therefore, we will disregard the complex solutions
because they are not good saddle points and of no physical relevance.
Hence for our model we use the physical real solution (20) to obtain:
$$M_{1, int.rigid}^2 =  -({\pi(D-2)\over 3})^2
{1\over \beta^2\epsilon}((1 -({\delta\over 1+\epsilon})^{1\over 2})^2 +
O(\delta,{\epsilon\delta\over (1+\epsilon)^2}))
\  .\eqno{(23)}$$
This is the main result of this letter. We now compare with the
large N-QCD result [5]:
$$M_{1,QCD}^2 = - {2NG_{QCD}^2\over \pi^2\beta^2}\eqno{(24)}$$
we obtain the following string solution for the QCD coupling constant:
$$NG_{QCD}^2 = ({4\pi^2 (D-2)\over 3})^2{1\over 3\kappa}\eqno{(25)}$$
where we have used (18) and neglected
$O(({\delta\over 1+\epsilon})^{1\over 2})$.  Using equation (15) we
derive from (25) the running beta function of the QCD coupling constant
we obtain:
$$\beta(NG_{QCD}^2) = -3({3\over 4\pi^2 (D-2)})^2[NG_{QCD}^2]^2\eqno{(26)}$$
producing correctly the asymptotic behaviour of QCD.  Modulo a constant
equation (26) has the correct structure of the QCD beta function.

In conclusion we have showed that the interacting rigid string with long
range K-R interactions gives a consistent solution for the free energy of
QCD at very high temperatures.  The solution gives a one-to-one
correspondence between the QCD coupling constant and the coupling constants
of the interacting rigid string theory.  Subsequently  we have proved that
the beta functions of the interacting rigid string theory to one loop
order not only it produces the correct structure of the QCD beta function
but it also produces the correct physics of asymptotic freedom.

\bigskip

{\bf Acknowledgement}

 I am very grateful to Profs. A. Polyakov and Y.Nambu for their constant
encouragement and extensive support over the last two years without which
we could not have gone far in our investigations. I also thank P.Ramond, and
F. Mansouri for constructive discussions and suggestions. Finally my
gratitudeto D. Zoller for fruitful collaborations over the years. I would like
to thankN.Khouri for his hospitality during my visit to the high energy
physics groupat Rockeffeler University. I also wish to thank J. Clark for his
assistance.

\bigskip

{\bf References}

\item {[1]} A. Polyakov, Nucl. Phys. B268 (1986) 406
; A. Polyakov,  Gauge fields, and Strings,
Vol.3, harwood academic publishers
\item {[2]} J.Polchinski and Z. Yang, Phys.Rev.D No.8 (1992) 3667.
\item {[3]} M. Awada and D. Zoller, Phys.Lett B325 (1994) 115.
\item {[4]} M. Awada, Cincinnati preprint Oct.(1994) to appear in
Phys. Lett B.
\item {[5]} J.Polchinski, Phys.Rev.Lett. 68, (1992) 1267.
\item {[6]} D. Kutasov, EFI-94-42  hep-th/9409128.
\item {[7]} D. Kutasov,  Nucl.Phys. B414 (1994) 33.
\item {[8]} M. Kalb, and P. Ramond
Phys. Rev. D Vol.9 (1974) 2237

\end